# Using single nanoparticle tracking obtained by nanophotonic force microscopy to simultaneously characterize nanoparticle size distribution and nanoparticle-surface interactions

Delyan R. Hristov[a], Dong Ye[a,†], Joao Medeiros de Araújo[a,b], Colby Ashcroft[c], Brian DiPaolo[c], Robert Hart[c], Christopher Earhart[c], Hender Lopez[a,*], Kenneth A. Dawson[a,*]

Comprehensive characterization of nanomaterials for medical applications is a challenging and complex task due to the multitude of parameters which need to be taken into consideration in a broad range of conditions. Routine methods such as dynamic light scattering or nanoparticle tracking analysis provide some insight into the physicochemical properties of particle dispersions. For nanomedicine applications the information they supply can be of limited use. For this reason, there is a need for new methodologies and instruments that can provide additional data on nanoparticle properties such as their interactions with surfaces. Nanophotonic force microscopy has been shown as a viable method for measuring the force between surfaces and individual particles in the nano-size range. Here we outline a further application of this technique to measure the size of single particles and based on these measurement build the distribution of a sample. We demonstrate its efficacy by comparing the size distribution obtained with nanophotonic force microscopy to established instruments, such as dynamic light scattering and differential centrifugal sedimentation. Our results were in good agreement to those observed with all other instruments. Furthermore, we demonstrate that the methodology developed in this work can be used to study complex particle mixtures and the surface alteration of materials. For all cases studied, we were able to obtain both the size and the interaction potential of the particles with a surface in a single measurement.

## Introduction

Nanotechnology for medical applications has been a topic of much scientific interest for several decades due to its potential to address existing challenges in patient treatment[1, 2]. In particular, the use of nanoparticles (NPs) as components in the design of targeted drug delivery has been an exciting concept. The field is now facing serious challenges partly due to the reduced circulation lifetime of NPs compared to more conventional approaches[3-5]. One of the main obstacles is that as NPs enter into living organisms they adsorb biomolecules forming a layer, known as the biomolecular corona. The NP-biomolecular complex has been shown to have a strong correlation with the "identity" of the materials and determine their biodistribution[6-8]. Mechanistic details of the interaction between this complex and cell/tissue surfaces in the body remain unclear, in part due to the lack of reliable methods to measure the physicochemical properties of materials in real exposure conditions. This includes, but is not limited to, accurate size of the particle-protein complex, NP-surface interaction at different conditions and the diffusivity of NPs close to a surface. Such information combined with other advanced characterization methods to study the accessible epitopes on the biomolecular corona could help elucidate the interaction mechanisms between NPs and the cell surface.

The most commonly property used to characterize a NP dispersion is its size distribution. For biological studies, the size of a NP is a crucial physicochemical property that affects its circulation in the bloodstream, its penetration into cells and tissues, and the activation of cell processes[9, 10]. However, it is not possible *a priori* to predict the biological activity, i.e. the toxicity, biodistribution, etc., of a material based solely on its size. There are other factors, such as the surface potential, which also have an effect. Thus different nanomaterials of the same size can lead to different final outcomes[11-14]. Nevertheless, it is widely recognized that for biological applications it is beneficial to have particles which operate in the same size range as most biological interactions (from a few nm to a few hundred nm)[15, 16]. For example, it has been reported that the size of NPs is a major factor that regulates the ability of NPs to penetrate into a poorly permeable tumors[17]. These observations highlight the relevance of an accurate size measurement of NP dispersion samples.

From a practical point of view, measuring the size of NPs is a relatively trivial task for simple materials, which may be accomplished through a plethora of techniques, such as dynamic and static light scattering (DLS and SLS)[18, 19], NP tracking analysis (NTA)[20], small angle X-ray scattering (SAXS)[21], differential centrifugal sedimentation (DCS)[18, 22], analytical ultracentrifugation (AUC), transmission electron microscopy (TEM)[18, 23] and many more. A major difficulty arises when measuring size distributions is to accurately resolve complex particles samples, i.e. composite materials or multicomponent dispersions. Light scattering techniques such as DLS and NTA are certainly a powerful tool for the analysis of these types of samples but in some circumstance may fail to resolve complex mixtures or not be applicable when used in realistic exposure concentration conditions[19]. Another widely used technique is DCS, which measures the precipitation time of particles under centrifugal force. If the particle density and shape are known the methodology can be relatively adept at resolving small size differences[22].

Though the size distribution of a sample is a fundamental measurable physicochemical property it is of limited use to understand and characterize the interaction of NPs with the cell surface. Certainly, the direct measurements of interaction forces between NPs and surfaces could shed light on the mechanisms that determine the final fate of NPs in living organisms. In the context of predicting the stability of colloid dispersion, NP-NP and NP-surface forces have been measured by methods such as colloidal probe atomic force microscopy[24-26] (AFM), total internal reflection microscopy[27-29] (TIRM) and more recently by nanophotonic force microscopy[30-32] (NFM). In a typical AFM experiment, a particle is immobilized on the tip of the cantilever and used as a probe to scan over other materials including surfaces, other immobilized particles or particles in solution. The main drawback of this technique is that the measurements have to be done at low temperature as the oscillations of the cantilever. Thermal fluctuations at room temperature are of the same order (or higher) as the forces due to NP-surface interactions[25, 33, 34]. On the other hand, TIRM and NFM are based on sampling the movement of NPs when they diffuse close to a surface. Then by analysis the changes in the distance between the NP and the surface, the force between them can be calculated[27, 30, 31]. Using these methodologies, forces of the order of 1 pN are resolved in experimental condition relevant for NP-cell interaction studies (in solvent and at room temperature). This method has also been employed to measure the size of polydisperse NPs samples[35]. TIRM has showed to be a versatile technique with a wide range of reported applications, including NP-surface potential interactions[27], steric repulsion[36], surface charge density[37], diffusion near a surface[38], Casimir force[39] and others. However, in all cases, the method is restricted to particles which are in the order of microns because gravity is the main force keeping them close to the surface. This experimental constraint, limits the use of TIRM to study NPs with possible applications in nanomedicine as these are, in most cases, only a few tens to a hundred nanometers in size. To address this size limitation, recently, the NFM has been proposed and used to measure NP-surface potential interactions[30, 31]. The main difference between these two methods is that in the NFM an optical trap is generated allowing for smaller size materials to be studied.

Naturally, a comprehensive particle analysis profits from multiple characterization methodologies. Furthermore, in the context of nanomedicine, the combined measurements of several physicochemical properties of a material (*e.g.* initial size and composition[40], shape, charge[40], surface functional groups[41], the presence, type and density of ligands[42, 43], possible impurities, etc.) are of interest because of their synergistic influence on NP "identity". We define such identity as a complex set of phenomena, such as the adsorption of molecules on the surface[6, 41] and change in interaction with biological entities due to the biomolecular corona or other physicochemical factors (e.g. Debye length)[40] which affect the biodistribution of materials. This intricacy of characterization combined with the need for detailed understanding of the particle dispersion presents a significant technical challenge to the community. Examples of new technologies used to address these issues are single particle optical extinction and scattering[44] and analytical ultracentrifugation[45].

Here we report a further development of the existing NFM technique to simultaneously measure the size distribution of a NP sample and its interaction potential with a surface. We apply the proposed method to analyze dispersions of bare silica NPs of 200 and 300 nm in diameter and to mixtures of these two sizes. We also study bare 200 nm silica particles and compare them to human serum albumin (HSA) coated and polyethylene glycol (PEG) grafted ones. We are able to not only obtain a high resolution measurement of the dispersion size distribution but also select and study selected subpopulations by changing the experimental parameters, such as laser power and solution salinity. Finally, we compare the size distribution obtained with our proposed NFM method to DLS, DCS and NTA in comparable conditions.

## Results and Discussion

### Single nanoparticle tracking

The first step to characterize a NP dispersion using NFM is tracking the trajectories of individual particles and then calculating average quantities or statistical descriptors based on the single measurements. For a detailed description of the bases of operation of NFM see reference [31]. Briefly the main components of the experimental set-up are: a waveguide (WG), a source of light (a laser) and a video camera used as a detector. Figure 1a shows a schematic representation of the NFM measurement chamber and sketches the forces acting on the NPs (see Figure S1 and S2 for images of the NFM used in this work and details on the chip configuration and the characterization of the WG). The basic operation of the NFM is as follows: the WG transports light from the laser through the experimental chamber which generates an exponentially decaying field that extends above the WG - referred to as the evanescent field. During an experiment, when a NP passes close to the WG it is trapped by the optical force generated by the gradient in the evanescent field ($F_{grad,z}$ and $F_{grad,y}$)[46]. As the trapped NP diffuses closer to the surface of the WG its motion is affected by the interaction force with the WG. At low solution ionic strength, this interaction is dominated by electrostatics repulsion ($F_{sur}$). The addition of these two forces perpendicular to the WG surface ($F_{grad,z}$ and $F_{sur}$) generates a potential well in the z direction. The functional form of the evanescent field in the z direction is well known[46] as a result by analyzing the time evolution of the intensity of the light scattered by a particle the fluctuations around a reference point in the z direction can be obtained (see Methods sections for more details). Particles also experience a force along the x axis (in the same direction as the propagation of light in the WG) due to the absorbance ($F_{abs}$) and scattering ($F_{scat}$) of the evanescent light. The movement of NPs in the x/y plane can be tracked with the scattered light observed by the camera (Figure S1). On the one hand, the force in the x direction propels the NPs over the WG and a displacement in this direction can be observed as shown in Figures 1c and S2a. On the other hand, the forces in the y direction confine the NP to stay preferably over the center of the WG as shown by the distribution of position shown in Figure S3b. By combing the direct tracking on the x/y plane with the z position obtained from the intensities analysis, a 3D trajectory of a NP can be reconstructed (Figure 1b). In particular, the fluctuations in the *z* direction can be used to calculate the interaction potential energy of the NPs with the surface of the WG and their size (see Methods section for details). Individual measurements can be coalesced into datasets and further used to calculate the size distribution of a sample. Certainly, the validity of the as described statistical descriptor will depend on the sampling. The methodology used here showed a high throughput, with a minimum of 200 NPs sampled per experiments which is comparable to NTA (>1000 NPs).

Additional potential benefits of the instrument can be found in its versatility, which can be coupled with size measurements proposed in this work. The NFM has been used for other applications not discussed here such as the measurement of diffusion coefficients of NPs travelling close to a surface[30] and for optical nanofluidic chromatography[46]. Finally, compared to other single NP tracking methodologies such as TIRM and Total Internal Reflection Fluorescence Microscope (TIRFM), NFM has two main advantages: it is able to trace sub-micron particles and it is label free.

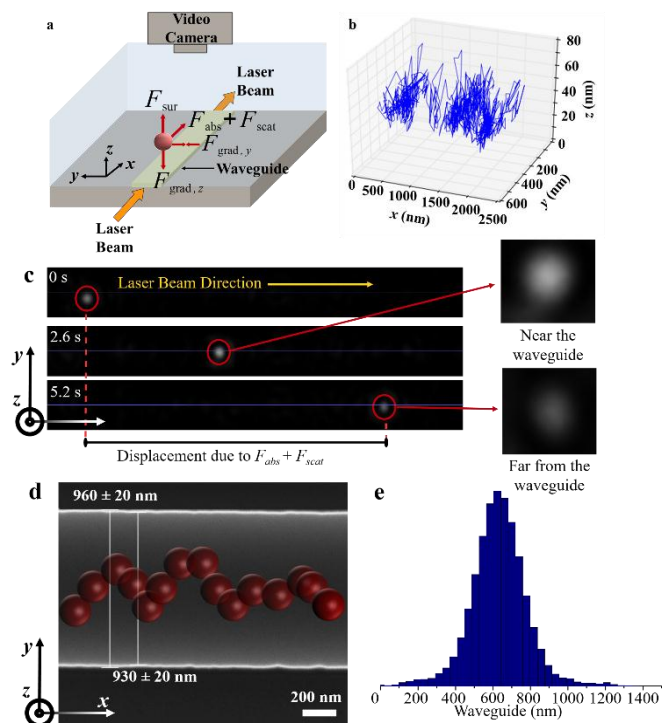

**Figure 1.** (a) Schematic representation of the experimental setup. (b) Example of a 3D trajectory of one NP. The data shown corresponds to the first 0.4 s of a trajectory that has a total duration of 4 s. (c) Images of a NP travelling over the WG at different times. Due the exponential decay of the evanescent field, NPs closer to the surface scatter more light and are detected brighter. The direction of the light in the WG is from left to right and so is the movement of the NP. (d) A schematic of a particle moving along the WG, to scale. (e) The distribution of the position of particles on the *y* direction of the WG for 200 nm silica NPs at $1.4 \times 10^{-3}$ M.

**NP-surface interaction**

To validate the methodology proposed we first study the interaction of Silica NPs with the surface of the WG under different ionic strength conditions. The methodology employed for the synthesis of all particles used in this work can be found in the Method Section and the SI. It is well known from standard colloid theory that in solution, NP-NP and NP-surface interactions are modulated by the concentration of ions in solution. As the ionic strength of the medium increases, the electrostatic repulsion decreases due to screening of the electrical double layer. In this work we accomplish this by increasing the concentration of PBS which predictably makes the NPs diffuse closer to the surface of the WG as depicted in Figure 2a.

Before analyzing the effect of the PBS concentration on the interaction forces, we present in Figure S4 the total potentials of a representative sample of particles which demonstrate the NP – to – NP variation in the potential and distance from the WG in a given sample. The ability of a precise measurement of the total interaction potential for single NPs will later be exploited to determine the size distribution of a sample. Additionally, each of these individual potential curves can be used to calculate averages quantities as shown in Figure 2b for the case of 200 nm Silica NPs in a PBS $1.4 \times 10^{-3}$ M. All average potential profiles are plotted with error bars (Figures 2b, 2c and 2d) which correspond to the standard deviations calculated for the average values. A qualitative inspection of the potential curves shows that the methodology captures the expected behavior, *i.e.* a potential well for the total potential and exponential relations for the optical and surface potentials. Also, the range of the energies (between -4 to 6 $k_B T$) is in agreement with previous reports of similar NPs in similar experimental conditions[30, 31].

The influence of the PBS concentration on the total potential energy is presented in Figure 2c which shows the expected behavior, *i.e.* an increase in ionic strength leads to a reduction in the estimate distance to the surface of the WG. Furthermore, the calculated values for the position of the minimums of the potentials well are in good agreement with the theoretical expected ones (Figure S5 and methods in SI).

The effect of the PBS concentration on the surface potential is shown in Figure 2d. We confirmed that our assumption that the interaction of the NPs with the surface of the WG is dominated by electric double layer repulsion, as the obtained potentials show the typical exponential decay predicted by Derjaguin, Landau, Verwey, and Overbeek (DLVO) theory. The Debye length was thus calculated from the fitting of the surface potential profiles and is compared to the expected values in Figure S6. We found that the measured and theoretical values were within 2 nm of each other which is similar to previous findings[22]. Notice that NFM overestimates the Debye length for all PBS concentrations. One possible source for this systematic error is the omission of the attractive interaction in the NP-surface potential[31]. Another possible factor is the assumption that the optical potential in the *y* direction is constant. In practice, the intensity a NP will reduce if it is closer to the edge of the WG even if its distance from the surface has remained constant. This will directly affect the calculation of the surface potential. In future works these two factors should be consider to improve the calculation of the Debye length. Despite this overestimation in all cases, the obtained Debye length show the trend of increasing with the ionic strength of the solution.

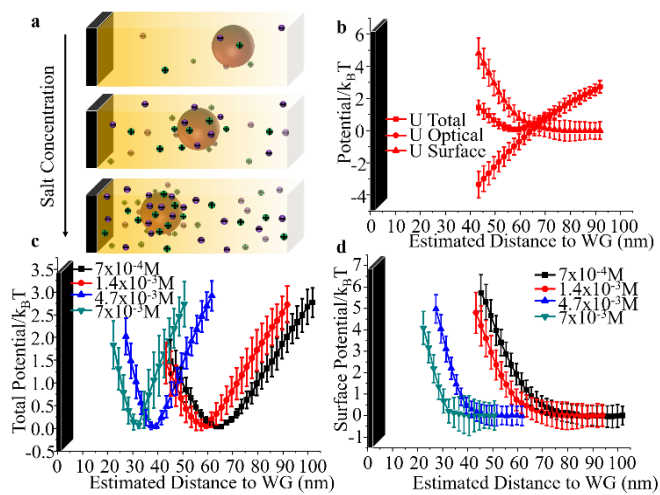

**Figure 2.** (a) A schematic representation of the effect of salt concentration on the position of a particle relative to the WG. (b) A representative example of the average total potential, optical potential and surface potential obtained for a single experiment (in this case 200 nm bare $SiO_2$ particles in I=1.4x10$^{-3}$M salt). (c) The total and (d) surface potentials of 200 nm bare $SiO_2$ particles with an increase in the salt concentration. Error bars correspond to the standard deviations of the average value.

**Particle size measurements**

As mentioned above, the potential well calculated for each NP can then be used to measure its size which may further be used to build a size histogram of the sample. The first step is to perform the optical fitting using Eq. 6. An example fit is presented in Figure 3a and shows that the proposed functional form for the fitting (Eq. 6) is adequate. Then, from the obtained fitting parameters the size of each individual NP can be obtained by the use of Eq. 8. At this point, all variables in this last expression are known except for the factor $I_0 e^{-\beta z_m}$ which is determined by a calibration procedure (see the Method section for details and Figure 3b for the curve used in this work).

The NFM method to measure size was compared to three standard characterizations techniques: DLS, DCS and NTA. We used 200 and 300 nm silica particles both bare and surface modified as models to study the capabilities of the instrument (Figure 3c and 3d and Table 1). DCS was chosen as a direct reference technique because it has been shown to be a very accurate way of measuring size, provided particle density and shape are known[22, 47].

As expected, the sizes observed by light scattering are marginally larger than the ones obtained by the other techniques. DLS and NTA measurements are based on particle diffusion hence the observed size corresponds to the hydrodynamic diameter. Furthermore, distributions obtained by DLS are the largest in observed size probably due to the disproportionate contribution of the biggest fraction of the particle population in an intensity type analysis as presented here[19, 48]. If the mean hydrodynamic diameter from DLS is calculated by number mean instead of intensity the result is much closer to the other instruments, 220 nm. The NFM and DCS data overlap and the FWHM (Full Width Half Maximum) procured from both instruments is very similar (Table 1). This strongly suggests that, once calibrated, size distributions acquired by NFM are not only accurate but also resolve size distributions only achievable by DCS which is considered as a high resolution method.

To further explore the applicability of the NFM we mixed and measured 200 and 300 nm bare silica particles (in a number ratio of 5 to 1) as a "representative" complex particle mixture. Additionally we studied the observed effect of altering the surface of the 200 nm particles by adsorbing human serum albumin (HSA) and separately by grafting polyethylene glycol (PEG).

**Table 1.** Mode sizes of 200 nm and 300 nm particles in NFM, DLS, DCS and NTA.

| NP diameter (nm) | NFM | FWHM | DLS | FWHM | DCS | FWHM | NTA | FWHM |
|---|---|---|---|---|---|---|---|---|
| 200 nm | 215 | 26 | 253 | 136 | 217 | 23 | 224 | 126 |
| 300 nm | 329 | 26 | 430 | 266 | 341 | 27 | 402 | 130 |

All observations were made at a solution ionic strength of 1.4x10$^{-3}$M and pH 7. Full-width half-max provided as a measure of the broadness of the observed peak.

Significant differences were observed between instruments when analyzing the mix of 200 nm and 300 nm particles. DLS, in both normal and high resolution mode, could not resolve the two populations and averaged over them (Figure 4a, dark gray line). The resulting broad peak had a mode value of 306 nm due to the signal contribution of the larger particles. The 300 nm fraction was observable in the NTA as a shoulder rather than a resolved population (Figure 4a, black dashes). As a result, the mode size value of the distribution was different than expected. However, the ratio between the two particle populations was in the correct range. DCS was able to resolve the two populations and the observed sizes were more similar to the expected values (Figure 4a, gray line). Both populations in relative ratio were observable in the NFM (Figure 4a, black line). The mode size strongly depends on the intensity of the evanescent field. At low power only the large particle population was trappable and could be studied independently. When the optical field intensity was increased smaller particles became observable (Figure 4b). In this way, by tuning the power of the laser, it was possible to separate populations from a distribution and study various sizes. This use of the instrument has previously been reported though in the context of optical chromatography[46].

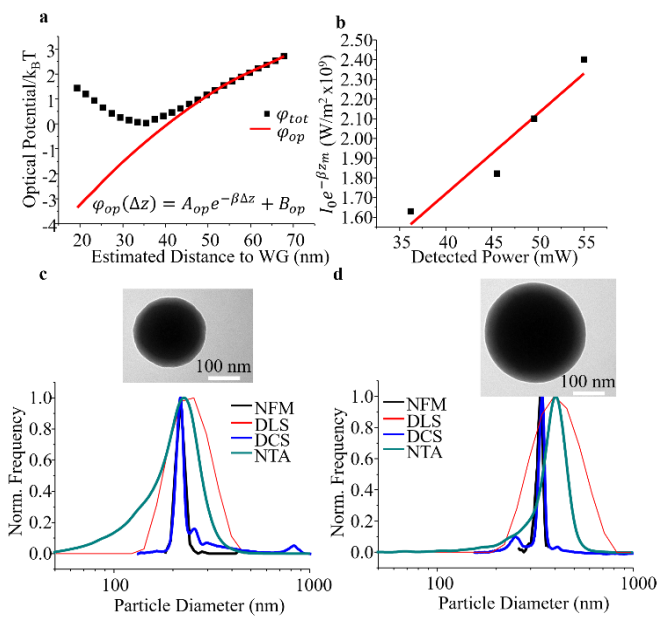

**Figure 3.** (a) A representative example of the fitted optical potential obtained from this experiment with the equation used. (b) A calibration curve obtained by manually fitting the results for 200 nm bare $SiO_2$ particles at different laser powers. $R^2$ for the linear fitting was 0.925 (c) TEM image and size distribution of 200 nm bare $SiO_2$ particles observed by NFM (independent measurement), DLS, DCS and NTA. (d) TEM image and size distribution of 300 nm bare $SiO_2$ particles observed by NFM, DLS, DCS and NTA. All observations were made at a solution ionic strength of 1.4x10$^{-3}$M and pH 7. Further TEM images and histograms are shown in Figure S7

When the HSA covered NPs were analyzed, all instruments showed a shift in the particle size distribution which could be interpreted as both protein adsorption and partial aggregation. The latter can be observed as an increase in the FWHM of the distribution in all methods. Only DCS was able resolve a mixture of HSA coated and bare silica particles (Table 2). A multimodal distribution for both HSA coated and mixed particles was observed by DCS (Figure 4c). This is likely due to differences in material density, i.e. particle populations with different amount of adsorbed protein. Because both density and size affect DCS measurements interpreting the data further is problematic. The NFM was unable to differentiate between populations with varying protein coverage DCS (Figure 4c). A 35 nm increase in size was observed when HSA, which is roughly 6.5 nm in diameter, was adsorbed on the particle surface. This inconsistency could partially be explained by the sample inhomogeneity (as seen by DCS) and observed minor aggregation post protein adsorption. It is possible that a much more thorough examination of such samples in various conditions could reveal multiple populations. The change in refractive index with the addition of protein could also affect the size measurements though this should be minimal due to the similarity of protein and silica electric permeability[49, 50]. Some of the error could be the result of the proximity of protein coated particles to the WG leading to unforeseen interactions (Figure 4c, surface potential insert). However, we think this is unlikely because there was no observable difference in the surface potential of the particles compared to others (insert in Figure 4c).

An increase in the particle size was observed in the NFM (22 nm), DLS (26 nm) and NTA (33 nm) post PEGylation (Figure 4d and Table 2). Considering the size of the ligand and its surface density we expect them to be partially extended, thus the measured change in size (about 10 – 15 nm a side) is reasonable. On the other hand, the reduction in particle size observed by DCS can be contributed to the change in particle density. In addition to the size characterization, using the NFM can also provide valuable information to understand the effect of coating the NPs on their behavior. For example, the insets in Figures 4c and 4d show that the addition of a protein or a PEG layer to the NPs leads to a diffusion closer to the surface of the WG. An effect which is most likely due to the reduction of particle surface charge. This assertion can be confirmed by observing that there is a correlation between average distance to the WG surface and change in zeta potential controlled by the degree of PEGylation (Figure S8).

Recently, it has been reported that NP hydrodynamic radius can be measured by NFM based on analyzing the diffusion of the particles as they are transported over the WG[35]. It is difficult to compare the two methodologies outright due to the differences in analysis conditions, including the material under investigation. Although the authors do not report size distributions we consider it possible that combining the methodology proposed here with a the diffusion based procedure outlined in [35] would lead to a more robust analysis procedure. Combining the optical and diffusional size could have additional unforeseen benefits for some studies, such as determining the electron permeability of single NPs. Thus in this way a more complete characterization procedure can be established.

On a practical level, the most user friendly instrument used here was the DLS. The process is mostly automated with comparatively few possible issues during measurements[19, 48]. DCS and NTA are available and usable by a capable professional, however there are some pitfalls. In the case of DCS it is the relationship between density, shape and centrifugation time which needs to be carefully considered for accurate measurements in complex conditions[47]. This was demonstrated with PEG grafted particles. On the flip side in some cases subpopulations which are difficult to measure with other techniques could be resolved in DCS due to this complex set parameters the instrument monitors (HSA adsorbed particles, Figure 4c). Newer NTA instruments are more user friendly and the software has become more automated, still some issues related to the concentration of particles used and thresholding remain[48]. In summary all three methods can be used with at a reasonable level with minimal training.

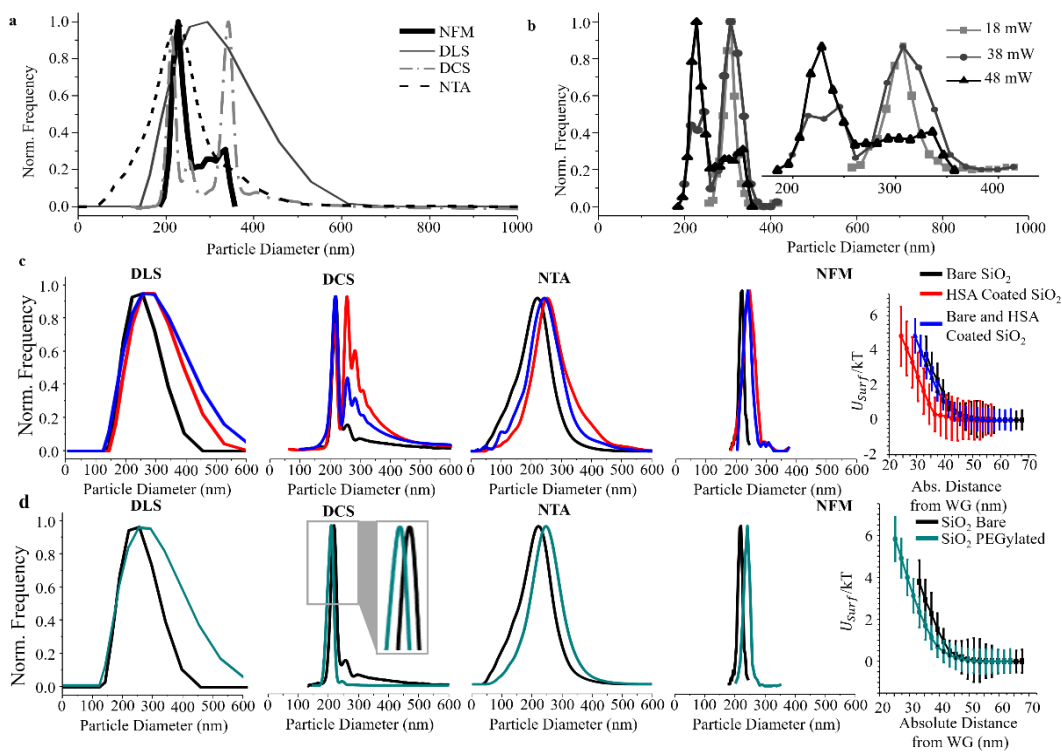

**Figure 4.** (a) The distribution of a mix of 200 and 300 nm particles observed by NFM, DLS, DCS and NTA. (b) The change in the observable particle subpopulation with increasing laser power from 18 mW where only the 300 nm population is measured to 48 mW where both populations can be observed. (c) Size distributions of bare and HSA coated 200 nm silica particles and a 1:1 (part./part.) mixture in (from left to right) DLS, DCS, NTA and NFM. The change in surface potential and average distance from the WG is also shown, furthest right. (d) Size distribution of bare and PEG coated 200 nm silica in (from left to right) DLS, DCS, NTA and NFM. The change in surface potential and average distance from the WG is also shown, furthest right. All observations were made at a solution ionic strength of $1.4 \times 10^{-3}$ M and pH 7.

|  | Laser Power (mW) | NFM | FWHM | DLS | FWHM | DCS | FWHM | NTA | FWHM |
|---|---|---|---|---|---|---|---|---|---|
| 200 nm | 38 | 217 | 15 | 253 | 116 | 217 | 23 | 213 | 113 |
| 200 nm and 300 nm | 18 | 305 | 25 |  |  | 213 /251 /340 | 22 /26 /29 | 221 /373 | 108 /83 |
| 200 nm and 300 nm | 40 | 231 /312 | 38 /41 | 306 | 171 |  |  |  |  |
| 200 nm and 300 nm | 48 | 227 /304 | 27 /81 |  |  |  |  |  |  |
| 200 nm HSA | 38 | 242 | 38 | 295 | 156 | 212 /256 /280 | 23 /24 /20 | 262 | 119 |
| 200 nm Bare and HSA Covered | 38 | 237 | 28 | 300 | 185 | 214 /257 /281 | 23 /14 /17 | 245 | 113 |
| 200 nm PEG | 38 | 239 | 20 | 279 | 122 | 207 | 24 | 246 | 105 |

Table 2. Mode sizes of bare 200 nm silica particles, a mixture of 200 and 300 nm particles HSA and PEG coated 200 nm particles and a mixture of bare and HSA coated particles as observed in NFM, DLS, DCS and NTA.

Each mode of multimodal dispersions is presented where relevant. Full-width half-max provided as a measure of the broadness of the observed peak. All observations were made at a solution ionic strength of $1.4 \times 10^{-3}$ M and pH 7.

In contrast, for the moment the NFM analysis remains difficult in terms of time and knowhow. An average measurement for one particle sample may take up to several hours for movie acquisition and a few additional hours for data analysis compared to 5 – 10 minutes for DLS and NTA, and 5 minutes to a few hours for DCS. The amount of sample required for NFM (1 - 10 μg/mL) is at least an order of magnitude lower than any of the other instruments used (DLS 10 - 100 μg/mL, NTA ~10 - 100 μg/mL, DCS ~100 – 1000 μg/mL). The lower size limit of particles which can be measured by NFM, though still dependent on material, is much higher than that of NTA, DCS and especially DLS. We found trapping 100 nm silica particles problematic and only possible at relatively high salt concentrations and laser power. As the trapping optical force depends on the dielectric constant of the material, metallic NPs of smaller size can be studied. It has been reported that Gold NPs as small as 20 nm can be trapped[51] with a similar WG and laser power the one used in this work. Characterizing the same dispersion was trivial in all other methodologies used. We expect that both ease of use and ability to analyze small particles will change as the instrument is further developed.

## Conclusions

To summarize we have outlined a method to successfully measure particle size of both simple and complex particle mixtures using NFM. We found that there is a good agreement between this methodology and techniques which are standard in the field of synthetic nanomedicine. Of special interest is the combination of size and surface potential measurements, and the possibility to separate particle populations by their surface properties. This experiments can be further diversified by coating the $Si_3N_4$ WG[52, 53] with anitbodies or proteins thereby providing more relevant information about the interaction of NPs with biological relevant surfaces. It is conceivable that with some modifications, the methodology could be coupled with an optical chromatography configuration. Overall the technique has unique benefits and some downsides compared to others presently in use, which suggests that it can be a part of a comprehensive analytical toolbox.

To an extent this study shows that in order to have a good understanding of a dispersion an appropriate choice of physicochemical parameters has to be monitored, especially when complex particles are the subject of investigation. This is highlighted when studying surface modified particles where the observations strongly depended on the method. Because of our combinatorial approach we know that there are several subpopulations of particles by protein surface density. However, the potential of these subpopulations and that of the bare particles cannot be differentiated near the surface. It is expected that the nature of the particle – WG interaction will become more divergent between coated and uncoated particles and possibly subpopulations with an increase of solution salinity. Especially at the transition point from diffusion over the WG to permanent adhesion. It is our opinion that this study showcases a possible application for our NFM methodology and how it may fit in the larger context of the field. Further it underlines how the synergetic use of several of these techniques can lead to a much more cohesive image of a particle dispersion, especially in complex conditions.

## Material and Methods

**Bare silica synthesis**

200 and 300 nm silica NPs were synthesized following a modified version of [54]. Briefly, 85 mL of methanol (Sigma Aldrich Prod. Code: 34860) were dosed from a bottle to the reaction flask. After which 25 mL of a 1 to 1 (V/V) methanol to ammonia (36%, Fisher Scientific Prod. Code: a/3280/pb17) and 7 mL of MiliQ water ware dosed into the same flask. The mixture was closed and left to equilibrate for 10 minutes. After the equilibration time 3.5 mL of Tetraethyl orthosilicate (TEOS, Sigma Aldrich Prod. Code: 333859) were then added to the synthesis mixture which was closed and left to react for an hour. The resulting 100 nm particles were allowed to undergo maturation for a further hour. The dispersion was diluted with the methanol ammonia solution made as described above by a factor of three and TEOS was pipetted into the unwashed dispersion at a rate of 1 mL/30 minutes until the particle size was measured to be 200 nm (characterized by DLS, number mean and DCS, relative weight). Some of the 200 nm particles were taken diluted again and regrown to 300 nm in the same way. It is important to keep the particle concentration relatively low to minimize aggregation.

The dispersion was spun at 4000 (3220 rcf) rpm for 20 minutes, the supernatant was replaced by MiliQ water. Particles were washed a total of four times. The final particle concentration was measured to be concentration of 65 mg/mL. Full details about this synthesis are available in the SI.

**Silica shell synthesis**

A secondary silica shell was added to the particles by adding 1% (V/V) of TEOS to the washed particle dispersion (10 mg/mL) at 90°C, stirring at 250 rpm similar to the procedure reported in [55]. The dispersion was left to grow overnight. Particles were washed four times as described above and stored at 25°C at a concentration of about 10 mg/mL.

**PEGylation of bare silica**

1 mL of 10 mg/mL 200 nm bare silica particles were spun (as above) and redispersed in MiliQ water before reaction. The dispersion was then heated to $90^0$C and allowed to equilibrate for 10 minutes while shaking at 1000 rpm. 5 kDa methylated PEG silicate (Iris Biotech GmbH Prod. Code: PEG4795) was added to the particles in a concentration of 10 ($2.7 \times 10^{-3}$M), 1 ($1.6 \times 10^{-4}$M) and 0.01 ($2.8 \times 10^{-6}$M) PEG/nm$^2$ to produce a range of surface densities. Particles produced this way are referred to as H, M and L respectively. PEG H: $7 \times 10^{-2}$ PEG/nm$^2$; PEG M: $4 \times 10^{-2}$ PEG/nm$^2$; PEG L: $3.1 \times 10^{-3}$ PEG/nm$^2$. The dispersions were left to react in this way for one hour after which they were washed four times in the same manner described above. Information on PEG density was done following the method described in [43]. Details are available in the SI.

**Preparation of HSA coated silica particles**

200 nm SiO$_2$ NPs (100 μg/mL) were incubated with human serum albumin (16.5 mg/mL) at 37 °C for 1 h with continuous shaking at 250 rpm. The NP-protein complex was pelleted from excess protein by centrifugation at 18 000 rcf, 4 °C for 1 min. The supernatant was discarded and the pellet was then resuspended in 1 mL MiliQ water and centrifuged again to pellet NP-protein complex. Particles were washed in this way a total of four times.

*Characterization Techniques*

To compare with the NFM method proposed here we employed three standard characterization techniques.

**Size distribution by Dynamic Light Scattering (DLS)**

A Malvern Zetasizer ZS series was used in all measurements. MiliQ water and solutions with the required salt concentrations were prepared and their conductivity and pH were measured by an Orion 3 Star Portable Conductometer and Benchtop pH meter respectively. Bare and PEGylated silica particles were diluted in the solutions by a factor of $10^3$ for a final concentration of ~100 μg/mL in a plastic low volume cuvette (PLASTIBRAND, semi-micro, PMMA, l = 1 cm). Particles were measured twice, both measurements consisted of two manual measurements each eleven runs for a total of forty four measurements. The number presented is an average of those measurements. ζ potential and surface ζ potential measurements can be found in the SI.

**Size distribution by Nanoparticle Tracking Analysis (NTA)**

A Malvern NanoSight LM 10 instrument was used for all measurements. The particles as measured in the DLS were taken from the cuvette and transferred into the NTA measurement chamber. Special care was taken to not have visible bubbles. Three 90 s movies were acquired and analyzed for all samples. In some cases camera exposure and movie threshold had to be readjusted for best results. The reported size and distribution is an average of those three measurements.

**Size distribution by Differential Centrifugal Sedimentation (DCS)**

DCS was performed using a CPS Disk Centrifuge DC 24000. 10 μL of clean particles at a concentration of 10 mg/mL were taken and dispersed in 90 μL of water or PBS (Sigma Aldrich Prod. Code: P4417) for a final concentration of 1 mg/mL. The disc speed of 18 500 rpm was used and an 8% - 24% water or PBS based sucrose (Medical Supply Prod. Code: 4821713) gradient was injected (settings optimized for size range analysis 0.03 – 1 μm). A 476 nm PVC commercial standard (Analytik UK) was used to calibrate the instrument before each measurement. Each gradient was checked by running the PVC standard as a sample and comparing to a database control. 100 μL of standard was injected before each measurement to calibrate the instrument.

**Size distribution by Transmission Electron Microscopy (TEM)**

Silica particles were diluted by a factor of 1000 with water and 10 μL were transferred on a Formvar carbon 200 mesh copper TEM grid (Agar Scientific) and left to dry in air overnight. The grid was imaged using FEI Tecnai 120 instrument using 120keV. Images were analysed using the ImageJ software.

**The NFM instrument**

A detailed explanation of the operation and physical principals of a NFM can be found in [31] while the full details of the instrument used and the settings are in the SI. Following we briefly describe the instrument used in this work, as well as the data processing and calculations performed.

Experiments were performed using the NT Surface system (Optofluidics Inc., Philadelphia). The experimental setup consists of a laser (635 mW, 1064 nm), a pneumatic pump to control the fluid flow (operation range from 0 to 70 mBar of pressure) and additional electronic and optics. The instrument is linked to a microchip mounted on a microscope stage and the microscope was further equipped with a camera (figure S1). Each NT Surface chip contains five silicon nitride ($Si_3N_4$) waveguides (WG): two 1, two 1.5 and one 2 μm wide which is situated in a 200 μm x 200 μm experimental chamber. Chips were provided by Optofluidics Inc. A 1064 nm laser light (TE mode polarized) is supplied by the instrument laser, coupled to the waveguides by the pre-aligned optical fibers, and guided to the waveguide outputs where optical power is measured with a photodiode. The intense scattering generated by particles enables high signal – to – noise imaging at low exposure time (100 μs) and high frame rates 2555 fps. The minimum time step that can be resolve will de determine by the frame rate employed. As reported in [35], with commercial cameras the maximum frame rate achievable is around 5000 fps which corresponds to a time step of 0.2 ms. Trapping objects on a waveguide were focused with an Olympus LUCPLFLN40X objective lens (0.6 NA) and images were captured for 20 seconds using a Basler acA2000-165uc camera. Images recorded by the camera were analyzed with a custom software package that performs automated particle tracking. Up to 30 movies were acquired for each sample, the specific number was varied as we attempted to keep the total number of particles relatively constant. In a typical experiment the length of the trajectories could vary from a few frames to up to around 35000 frames (corresponding to approximately 13 s). Only trajectories with at least 3000 frames (1.1 s) where considered for the calculations explained below.

**Sample Preparation for the NFM**

200 nm and 300 nm silica NPs bare, with surface grafting of various densities of PEG or coated with proteins were diluted with PBS in water (Sigma Aldrich Prod. Code: 34877) at the appropriate concentration by a dilution factor of about $10^4$. This resulted in a final concentration of $10^7$ particles per mL for all samples.

**Movie Analysis**

Movies were analysed using a custom Trackmate based software in Fuji (http://fiji.sc/Optofluidics) developed by Optofluidics Inc. To ensure an adequate statistical sample, for all calculations, only trajectories with at least 3000 frames were used. More details about the settings used and some additional considerations are available in the SI.
To correct a systematic drift of the observed in time sequences of the intensity measurements a high pass Butterworth digital filter was applied. Care was taken to verify that the filtering process did not affect the potential energy calculations.

**Calculation of the total and surface potential from the NFM trajectories**

From the movie analysis, the time evolution of the position in the $x-y$ plane and the intensity of the light scattered for each NP tracked was obtained (see Figure 1a for the definition of the coordinate system). The intensities can be used to study the movement of the NP in the $z$ direction (perpendicular to the WG) as it is known that the WG generates an exponentially decaying field that extends above its surface. Then, the fluctuations of the position in the $z$ direction are used to calculate the interaction potential between the NP and the surface of the WG. A similar principal is used in TIRM. Following we briefly review how to calculate the total and surface potential from the intensity measurements. For a detailed justification of the calculations shown here, we refer the reader to [31, 32].

As previously mentioned, it is well known that the evanescent field decays exponentially, this means that the intensity of light ($I$) scattered by a NP will also decays exponentially as a function of the distance to the WG ($z$):

$$I(z) = I_0 e^{-\beta z}, \quad (1)$$

where $I_0$ is the intensity measured for a NP that is in contact with the WG ($z = 0$) and $\beta$ is the inverse of the penetration depth of the evanescent field. If $I_0$ is known, directly from Eq. 1 the intensity data could be transformed into distance. Then this data is used to calculate the probability of finding the NP at a distance $z$ for the surface which can, in turn, be mapped into the potential energy interaction between the WG surface and the NP (assuming a Boltzmann statistics). However, the value of $I_0$ is in most cases unknown, so in practice the set of intensities of a single NP is used to build a histogram. The intensity with highest frequency, $I(z_m)$, is used as reference (this will correspond to the equilibrium separation, $z_m$, for the interaction between the NP and the WG). Therefore, the total potential energy for the interaction between the NP and the WG surface as a function of the relative distance between, $\Delta z = z - z_m$, is given by:

$$\frac{\varphi(\Delta z)}{k_B T} = \frac{\varphi(z) - \varphi(z_m)}{k_B T} = \ln \frac{N[I(z_m)] \, I(z_m)}{N[I(z)] \, I(z)}, \quad (2)$$

where $\varphi(z_m)$ is the minimum of the total potential energy, $N[I(z)]$ the number of observations of intensity $I$, $N[I(z_m)]$ the number of observations of intensities $I(z_m)$, $k_B$ is the Boltzmann constant and $T$ is temperature. The relative distance, $\Delta z$, is given by:

$$\Delta z = z - z_m = -\frac{1}{\beta} \ln\left(\frac{I(z)}{I(z_m)}\right), \quad (3)$$

The calculated total potential as explained above is composed of two main contributions: the optical trapping potential that pushed the NP toward the WG, and the interaction between the surface of the WG and the NP which for the conditions chosen in this work (type of NP, material of the WG, salt concentration) is mainly electrostatic repulsive. The combined effect of the two opposite forces creates a potential well as illustrated in Figure 2b and Figure 2c. In this way the NP-surface interaction potential, $\varphi$, can be obtained by:

$$\varphi_s(\Delta z) = \varphi(\Delta z) - \varphi_{op}(\Delta z) + \varphi_s(z_m) + \varphi_{op}(z_m), \quad (4)$$

where is $\varphi_{op}$ is the optical trapping potential, $\varphi_{op}(z_m)$ is the optical potential at $z_m$, and $\varphi_s(z_m)$ is the surface potential at $z_m$. Notice that in Eq. 4 the individual contributions to the total potential are written as a function of $\Delta z$ as this is the argument of the total potential obtained from the measurements and that the contribution from the reference potential are also included. For particles which are smaller than the wavelength of the incident light (Rayleigh regimen), the functional form of $\varphi_{op}$ has been established from theoretical studies[46] and also confirmed by numerical simulations[56] and in term of $\Delta z$ is given by:

$$\varphi_{op}(\Delta z) = \frac{2\pi}{c} \alpha I_0 e^{-\beta z_m} e^{-\beta \Delta z}, \quad (5)$$

where $\alpha$ is the polarizability of the particle, $\beta$ is the inverse of the permeability of the evanescent field and $c$ the speed of light. With the known functional form of the optical trapping potential, in practice $\varphi_{op}$ can be obtained by numerically fitting the total potential to an exponential of the form of Eq. 5 in a region far from the equilibrium position as in this region the contribution of $\varphi_s$ to the total potential is negligible[32]. In practice the fitting is performed to the following expression:

$$\varphi_{op}(\Delta z) = A_{op} e^{-\beta \Delta z} + B_{op}, \quad (6)$$

where we fix the penetration depth ($1/\beta$) to 57 nm as this is a known parameter of the waveguide[32] and $A_{op}$ and $B_{op}$ are fitting parameters. The fitting will also account for the reference potential in Eq. 4. After $\varphi_{op}$ is calculated, $\varphi_s$ is obtained by $\varphi_s = \varphi - \varphi_{op}$. It is important to highlight at this point that the main assumption for this procedure was that the optical potential (which is obtained by the fitting process) dominates at distance far from the surface. This will certainly be the cases in most experimental conditions as the penetration depth of the evanescent field is grater (around 60 nm) than the Debye length for a typical solution in which the experiments are performed (less than 12 nm).

*Obtaining the Size from NFM*

The fitted $\varphi_{op}$ can also be used to calculate the size of the NP. The polarizability of the particle depends on the size by:

$$\alpha = \frac{4\pi R^3 (\varepsilon_p - \varepsilon_m)}{\varepsilon_p - 2\varepsilon_m}, \quad (7)$$

where $\varepsilon_p$ and $\varepsilon_m$ are the relative permittivity of the particle and the medium and $R$ is the radius of the NP. Comparing Eq. 5 with Eq. 6 and using the definition in Eq. 7, we have that the prefactor of the exponential obtained from the fitting of the optical potential ($A_{op}$ in Eq. 6) can be used to determine the size of the NP by:

$$R^3 = \frac{A_{op} c \, e^{\beta z_m}}{8\pi^2 I_0} \frac{(\varepsilon_p - \varepsilon_m)}{\varepsilon_p - 2\varepsilon_m}, \quad (8)$$

In the expression above $\beta$ is a known parameter of the WG. If the material from which the NPs are made is known then $\varepsilon_p$ and $\varepsilon_m$ are also known parameters. Silica can be considered a materials with low optical absorption we assume $\varepsilon_p \approx n_p^2$ where $n_p$ is the refractive index of the particle. The same approximation is done for the medium, i.e. $\varepsilon_m \approx n_m^2$. The values of $n_p$ and $n_m$ used where 1.45 and 1.33 respectively. This leaves $I_0$ and $z_m$ as unknown parameters in Eq. 8 which for a given material and WG geometry will mainly depend on the power of the laser ($P$). A relatively straightforward procedure to obtain $I_0$ is by sticking a NP to the surface and recording its intensity. This is commonly used in TIRM experiments but for our experimental setup this proved impractical. Furthermore, even if $I_0$ was measured the dependence of $z_m$ on the laser power is also unknown. For these reasons, we employed a calibration procedure to determine the factor $I_0 e^{-\beta z_m}$, referred to later as the calibration factor. A dispersion of 200 nm Silica NPs previously characterized by DCS was measured at different laser powers. At each laser power, $I_0 e^{-\beta z_m}$ was adjust to so that the maximum of the NFM and DCS size distributions match. The obtained calibration curve is reported in Figure 3b in which a linear fit was performed. A priori it is known that $I_0$ depends linearly on $P$, but the dependence of $z_m$ on the evanescent field is unknown. Calculating $z_m$ (outlined below) for the set of laser powers used for the calibration, we find a linear dependence (Figure S8) with a slope which is 2 orders of magnitude lower than $\beta$. This means that the term $e^{-\beta z_m}$ can be considered constant in the range of $P$ used in this work. Thus our assumption that $I_0 e^{-\beta z_m}$ is approximately linearly dependent on $P$ is justified.

**Estimation of the distance to the surface of the WG**

To estimate the equilibrium distance to the surface of the WG, $z_m$, we first recognize that for the experimental conditions used in this work (PBS $7 \times 10^{-4}$ M to $7 \times 10^{-3}$ M, corresponding to Debye lengths of 11.5 and 3.6 nm) the contribution of van der Waals interactions between the NP and the WG surface is negligible compared to electrostatic interactions. With this assumption, and using Derjaguin, Landau, Verwey, and Overbeek (DLVO) theory we have that $\varphi_s$ can be approximated by [57]:

$$\varphi_s = 16\varepsilon_m R \left(\frac{k_B T}{e}\right)^2 \tanh\left(\frac{e\psi_S}{4k_B T}\right) \tanh\left(\frac{e\psi_p}{4k_B T}\right) e^{-\frac{z}{\lambda_D}} \quad (9)$$
$$= k_{el} e^{-\frac{z}{\lambda_D}},$$

where $e$ is the fundamental charge of the electron, $\lambda_D$ the Debye length, and $\psi_S$ and $\psi_p$ are the Stern potentials of the surface and the particle, respectively. From the experimental procedure described above, the $\varphi_s$ is calculated and fitted to $A_{el} e^{-\Delta z/C_{el}}$. Then we compare the parameter from the fitting to the theoretical prediction (Eq. 9):

$$A_{el} = k_{el} e^{-\frac{z_m}{C_{el}}}. \quad (10)$$

The factor $e^{-z_m/C_{el}}$ is a consequence of $\varphi_s$ being measured as a function of $\Delta z$ and not of $z$. Eq. 10 can be solved for $z_m$ if $k_{el}$ is estimated (see definition of $k_{el}$ in Eq. 9). $R$ has been calculated for each NP while $\varepsilon_m$, $k_B T$ and $\lambda_D$ are parameter that can be determined from the experimental conditions. All measurement were done at room temperature giving $k_B T = 4 \times 10^{-21}$ J. For the PBS solutions used $7 \times 10^{-4}$, $1.4 \times 10^{-3}$, $4.67 \times 10^{-3}$ and $7 \times 10^{-3}$ M $\lambda_D$ correspond to 11.5, 8.1, 4.4 and 3.6 nm respectively. For $\varepsilon_m$ we have $710 \times 10^{-12}$ C$^2$N$^{-1}$m$^{-2}$. The only parameters missing to evaluate $k_{el}$, are the Stern potentials. As a direct measurement is not possible, we instead use the measured value (DLS) for the zeta potential of the particles and the WG (details in SI) to replace the Stern potentials in Eq. 9. Finally, notice that parameter $C_{el}$ is the fitted Debye length which can also be used to validate the proposed methodology (see main text).

## Acknowledgements

D.R.H would like to acknowledge the EU FP7 project FutureNanoNeeds (Grant agreement no: 604602). D.Y would like to acknowledge Enterprise Partnership Scheme Postdoctoral Fellowship Programme (Project ID EPSPD/2014/5). J.M.A. acknowledges CNPq - Conselho Nacional de Desenvolvimento Científico e Tecnológico. R.H., C.A., B.D. and C.E. would like to acknowledge US National Science Foundation. H.L would like to acknowledge the financial support of the Irish Research Council, Enterprise Partnership Scheme Postdoctoral Fellowship Programme (Project ID EPSPD/2015/5). The authors would like to acknowledge Dr. Bernardo Cordovez, Dr. Sergio Anguissola and Dr. Denis Headon for their contribution at the early stages of the project.

**Supporting information for: "Using single nanoparticle tracking obtained by nanophotonic force microscopy to simultaneously characterize nanoparticle size distribution and nanoparticle-surface interactions"**

*Delyan R. Hristov[a], Dong Ye[a] , Joao Medeiros de Araújo[a,b], Colby Ashcroft[c], Brian DiPaolo[c], Robert Hart[c], Christopher Earhart[c], Hender Lopez[a,]\*, Kenneth A. Dawson[a,]\**

[a] Center for BioNano Interaction, School of Chemistry, University College Dublin, Belfield, Dublin, Ireland

[b] Departamento de Física, Universidade Federal do Rio Grande do Norte, Natal-RN, Brazil

[c] Optofluidics, Inc., Philadelphia, PA 19104 USA

\*E-mail: Hender.LopezSilva@cbni.ucd.ie

\*E-mail: Kenneth.A.Dawson@cbni.ucd.ie

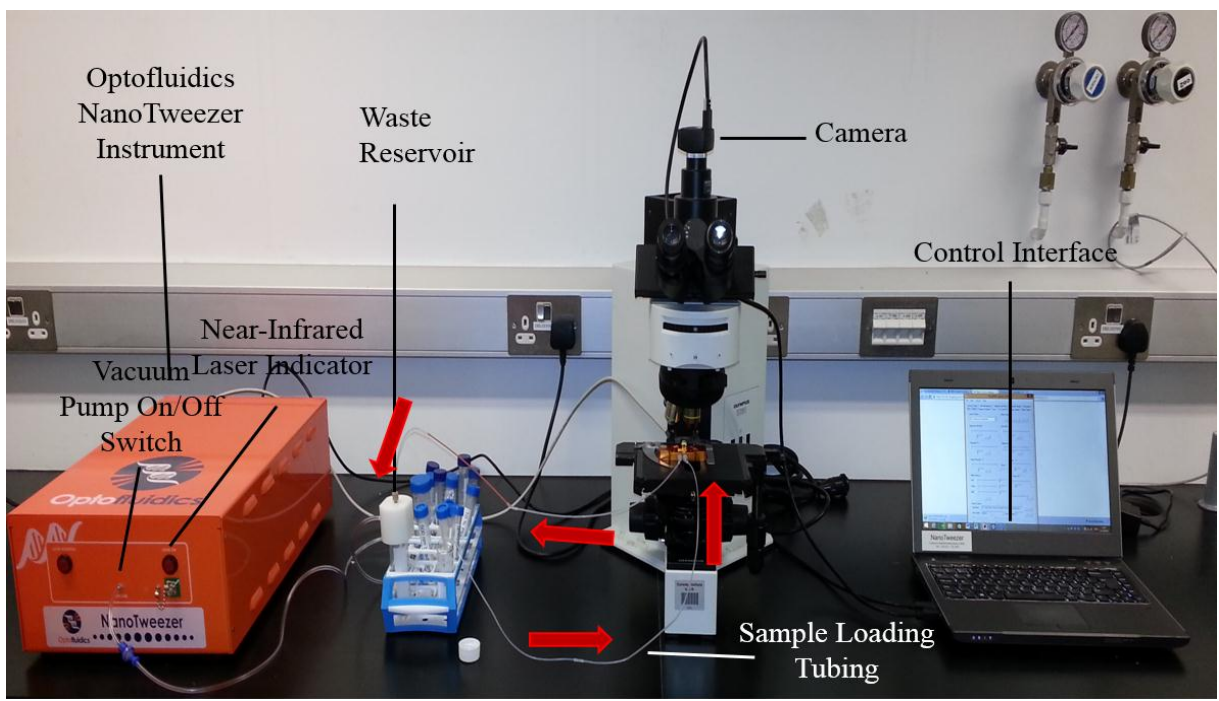

**Figure S1.** Picture of the NFM instrument setup. The red lines show the flow of liquid from the sample to the waste chamber.

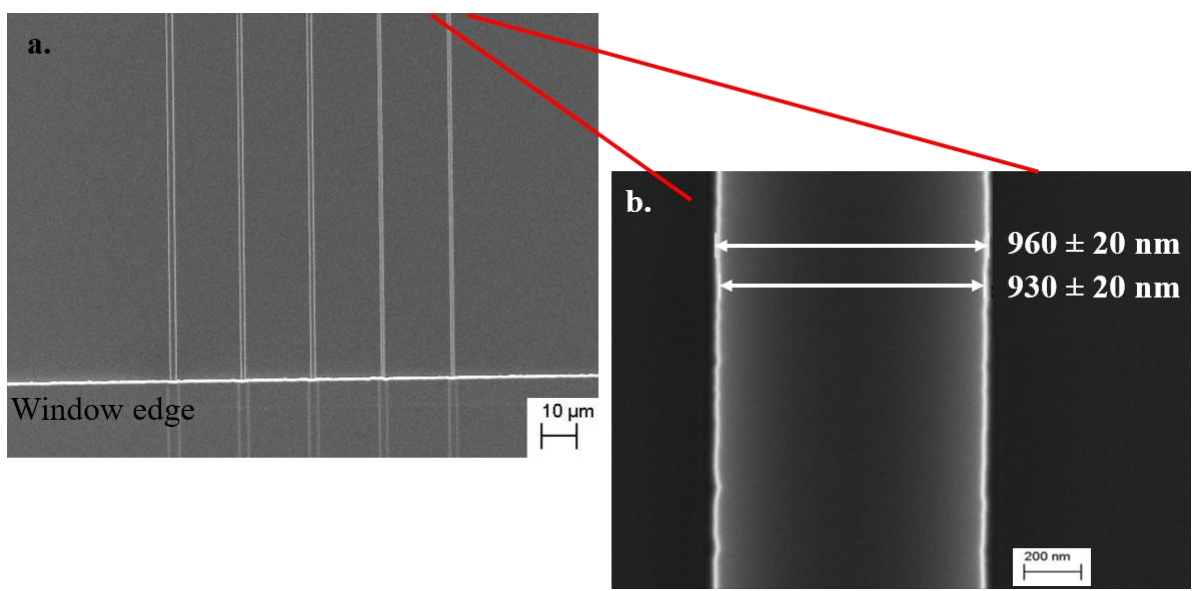

**Figure S2.** Anatomy of a standard measurement window. (a) Shows a view of a large section of the measurement window while (b) is an SEM close-up of the edge of the experimental window and a single WG, the same size as the ones used in the study.

S - 2 -

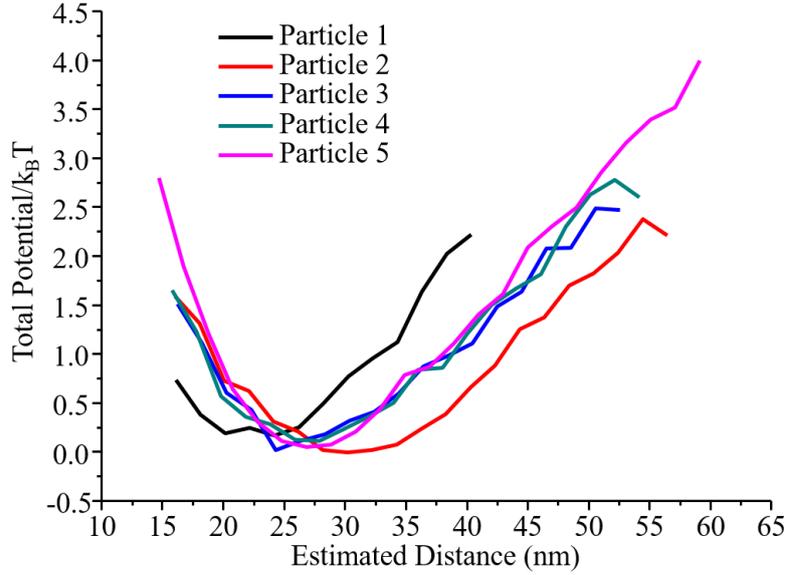

**Figure S3.** Example of the variability of the total potential energy and the distance from the WG surface for five representative individual particles. Sample analyzed is 200 nm silica at $1.4 \times 10^{-3}$ M.

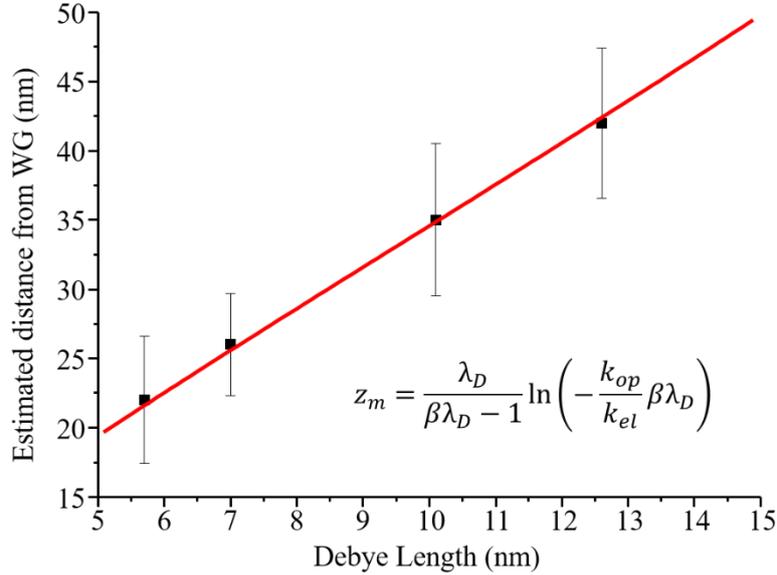

$$z_m = \frac{\lambda_D}{\beta\lambda_D - 1} \ln\left(-\frac{k_{op}}{k_{el}}\beta\lambda_D\right)$$

**Figure S4.** Estimated distance to the WG as a function to the Debye length. The continuous line is a fit to the theoretically predicted expression shown in the figure (see equation S2 and the methods section in the SI for more details). Error bars correspond to the standard deviations of the average estimated distance to the WG.

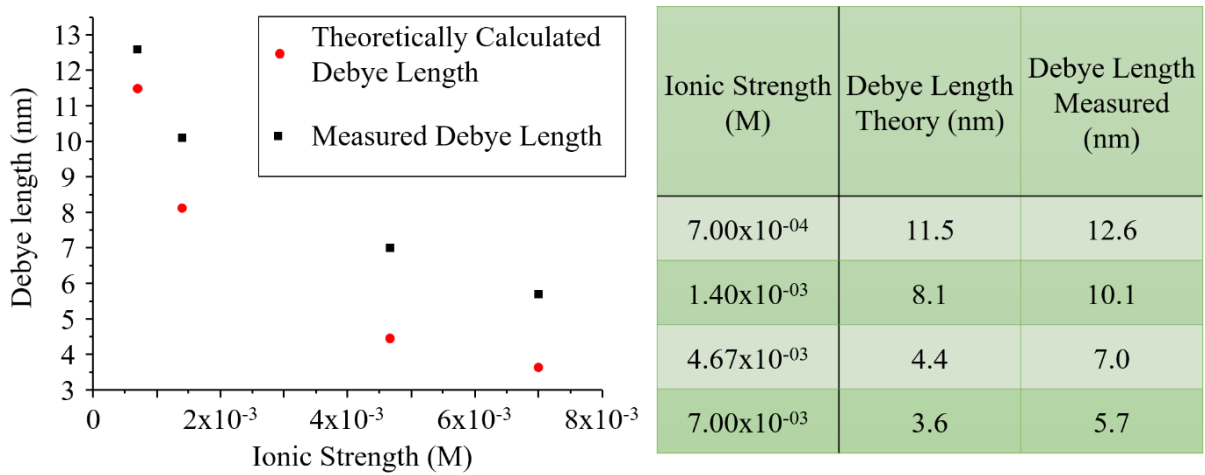

**Figure S5.** Debye lengths measured by NFM (Eq. 10 in main text) compared to theoretical values.

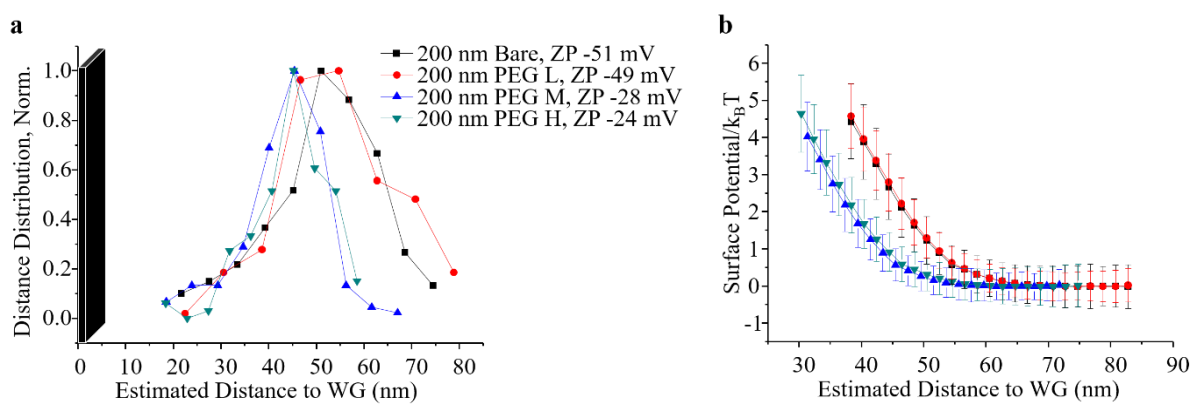

**Figure S6.** (a) Distance distribution of a particle dispersion with varying PEG surface concentration (PEG H: $7 \times 10^{-2}$ PEG/nm$^2$; PEG M: $4 \times 10^{-2}$ PEG/nm$^2$; PEG L: $3 \times 10^{-3}$ PEG/nm$^2$) and (b) surface potentials of the same particles. All observations were made at a solution ionic strength of $1.4 \times 10^{-3}$ M and pH 7. Error bars correspond to the standard deviations of the average value.

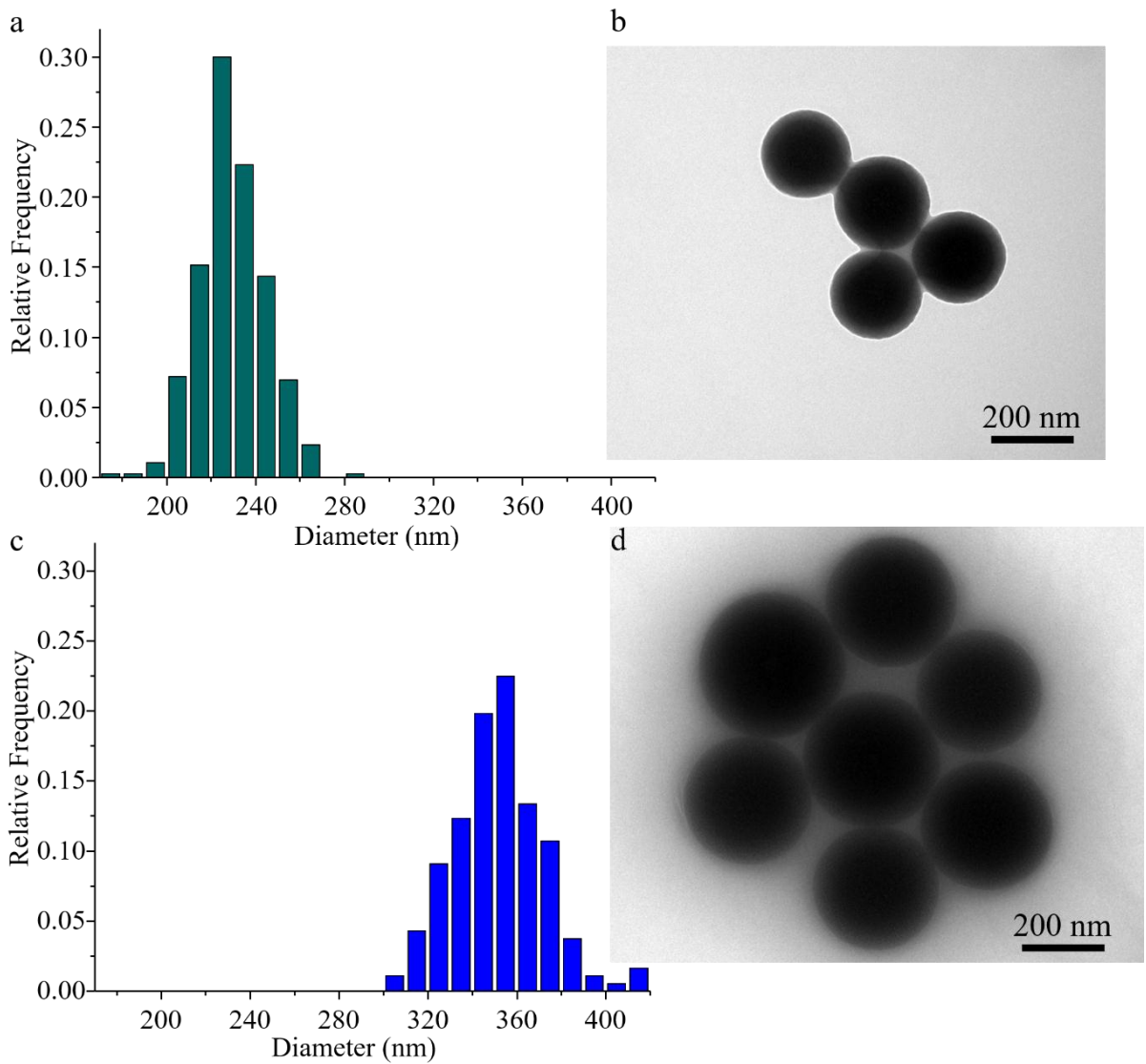

**Figure S7.** TEM analysis of bare SiO$_2$ particles used in this study. (a) Histogram and (b) representative TEM image of 200 nm SiO$_2$ particles. (c) Histogram and (d) representative TEM image of 200 nm SiO$_2$ particles.

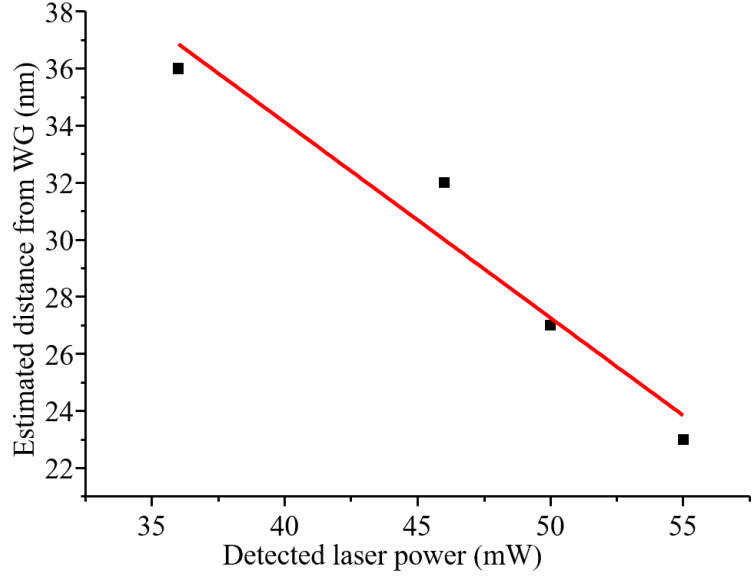

**Figure S8.** Changes in average particle distance with increase of detected laser power. The continuous line is a linear fit to the data.

**Dependence of the position of the minimum of the total potential energy on the Debye length**

As mentioned in the method section in the main text, the total potential energy for the interaction of a NP with the surface of the WG can be approximated by the sum of two main contributions: the optical potential generated by the evanescent field ($\varphi_{op}$) and the electrostatic repulsion from the electric double layer ($\varphi_S$). The complete expression for these two contributions are given in Eq. 5 and Eq. 9 in the main text. Taking the derivative of the total potential energy with respect to $z$, we have:

$$\frac{d\varphi(z)}{dz} = -\frac{k_{el}}{\lambda_D} e^{-\frac{z}{\lambda_D}} - k_{op}\beta e^{-\beta z}, \tag{S1}$$

with $k_{el}$ as defined in the main text and $k_{op} = 2\pi\alpha I_0/c$. Expression S1 is zero at the position of the minimum ($z_m$) and solving for $z_m$ we obtain:

$$z_m = \frac{\lambda_D}{\beta \lambda_D - 1} \ln\left(-\frac{k_{op}}{k_{el}} \beta \lambda_D\right), \tag{S2}$$

which gives the dependence of $z_m$ with the Debye length. Notice that we have neglected the dependence of $k_{el}$ on $\lambda_D$. This is justified as $k_{el}$ depends as the tanh of the Stern potentials.

**Synthesis**

*Bare silica synthesis*

Before use a jacketed three necked flask was rinsed with methanol and left to dry. Then it was placed on a magnetic stirred and connected to a water bath with a build in pump (temperature is set to 40°C). The flask was supplied with a thermometer (neck 1), a stopper (neck 2) and another stopper (neck 3) and a 25 mm egg shaped rare earth magnetic stirrer. Typically neck 2 was used for addition of compounds during the reaction.

A Syrris Atlas asp - 030 syringe pump was used to add all the reagents to the flask. The pump and tubing were rinsed with methanol. 85 mL of methanol were dosed from a bottle to the flask (intake though port A and discharge though port B where the tube was put though neck 2).

After the dosing was complete the liquid was removed by pumping though with air in a waste bottle for two syringe fulls (5 mL). In the meantime 35 mL of a 1 to 1 (V/V) methanol to ammonia (36%) solution was prepared. The syringe was equilibrated with the methanol/ammonia solution for two syringe volumes (5 mL) and then 25 mL of the solution was dosed into the flask in the way described above.

The pump and tubing were again dried by running air though them and rinsed first with methanol for two cycles and then with water for another two. 7 mL of water was then dosed into the flask after which the system was closed and left to equilibrate for 10 minutes. At this time the temperature was measured to be 40°C.

Port C was rinsed using methanol followed by air for two syringe cycles of both. The pump was equilibrated with TEOS by flushing for one syringe volume. 3.5 mL of TEOS were added to the synthesis mixture. Following this neck 2 was closed and the mixture was left to react for an hour followed by a maturation for a further hour. The dispersion was diluted with the methanol ammonia solution made as described above by a factor of three and TEOS was pipetted into the unwashed dispersion at a rate of 1 mL/30 minutes until the particle size was measured to be 200 nm (characterized by DLS, number mean and DCS, relative weight). Some of the 200 nm particles were taken diluted again and regrown to 300 nm in the same way. It is important to keep the particle concentration relatively low to minimize aggregation.

The dispersion was spun at 4000 (3220 rcf) rpm for 20 minutes, the supernatant was replaced by MiliQ water. Particles were washed a total of four times. The final particle concentration was measured to be concentration of 65 mg/mL

**Characterization of bare, HSA covered and PEGylated silica particles**

*ζ potential by Zetasizer ZS*

After the size measurement, particles were transferred to a zeta potential cuvette (DTS1070) and measured three times using manual measurements with eleven runs each for a total of thirty three runs. The zeta potential presented is an average of the final values. All measurements used the Smoluchowski model. The conductivity obtained from the Zetasizer instrument was always above 0.2 mS/cm. pH of all dispersant solutions was between 7 and 7.4.

*Surface of the WG zeta potential measurement*

The surface zeta potential was measured following the protocol provided by the company[1]. Briefly, the $Si_3N_4$ surface, provided in the appropriate dimensions by Optofluidics Inc. was glued to a holder and left to settle. The holder was inserted in a ZEN1020 cell. Before the measurement was carried out the zero distance of the surface was found [1-2]. The surface zeta potential was measured using 200 nm bare silica particles at a concentration of 100 μg/mL dispersed in a solution with an ionic strength of $1.4 \times 10^{-3}$, $2.8 \times 10^{-3}$ and $1.4 \times 10^{-2}$ M. The measured zeta potential was -77 mV (n=2) for $1.4 \times 10^{-3}$, -71 mV (n=2) for $2.8 \times 10^{-3}$ and -67 mV (n=1) for $1.4 \times 10^{-2}$. The pH of all solutions was around 7.2. These results are similar to the ones described in [3].

**PEG density analysis**

*Particle dissolution studies in DLS (data not shown)*

Core silica particles were placed in a plastic cuvette (l = 1 cm) at a concentration of 10 mg/mL and NaOH was added so that the final concentration was 50 mM, 100 mM and 200 mM. The cuvette was sealed well with parafilm and placed in the instrument where a size measurement was taken every 5 minutes for twelve hours (temperature was set to $37^0$C) with a fixed attenuator of 11. The drop in count rate over time was observed. In this way the standard particle dissolution procedure was established. Particles were dispersed in 200 mM NaOH in a 2 mL Safe Lock® Eppendorf and left 16 hrs at $37^0$C. Particle dissolution is then confirmed using a standard DLS measurement.

*Nuclear Magnetic Resonance*

In all cases a 5 mm thin wall, 8 inch, 500 MHz NMR tube was used (Wilmad Lab Glass). Oxford instruments 400MHz was used for all measurements. Samples were measured with a minimum of 16 scans and a relaxation time of 25 s. No changes in the angle or temperature were made to the default protocol ($25^0$C with no equilibration time and $45^0$ detection angle).

*Silicate PEG5000 OMe calibration curves*

5 kDa methylated PEG silicate at a concentration of 5.7 mg/mL (1.1 mM), 2.8 mg/mL ($5.7 \times 10^{-1}$ mM), 1.4 mg/mL ($2.8 \times 10^{-1}$ mM), 0.7 mg/mL ($1.4 \times 10^{-1}$ mM), 0.35 mg/mL ($7 \times 10^{-2}$ mM) was dissolved a 1 mL solution of 200 mM NaOD with 1mM DMF as an internal standard. The solution was left overnight at $37^0$C to match dissolution conditions and then measured. Relaxation time in the measurement was 25 seconds and each measurement consisted of 32 scans. Two calibration curves were averaged. The mean PEG peak (3.6 ppm) was used to determine concentrations for the calibration curve.

*Dissolution $^1$H NMR*

An aliquot of NaOD (5M in D2O) was added to a known concentration of particles, typically between 3 and 10 mg/mL, so that the final base concentration is 0.2 M. The dissolution procedure was as described above i.e. incubation for ~16 hrs at 37$^0$C

The dissolved particles were put in a clean NMR tube and measured in a 400 MHz NMR (16 scans and 25 seconds relaxation time).

*NMR spectra Processsing*

MestReNova 8.0 software was used for peak fitting to determine the integrated area and FWHM. All NMR spectra were processed in the following fashon: the obtained spectra was chemical shift referenced using the internal standard (DMF) and standard solvent peak, an exponential apodization of 0.3 Hz was then applied, the phase of the spectra was corrected manually (if required). Spectra were baselined using Polynomial Fit in all cases. Peak picking and integration and peak fitting were done in manual mode using the MNova software and the areas were compared to the calibration line described above.

*Calculations of PEG density*

The following assumptions were made in all calculations used in the work: all particles are perfect spheres of the same size, the density of all particles is the same and equal to 2 g/cm$^3$. DCS was assumed to give an accurate description of "true size" for bare particles and is used for further calculation. To obtain the concentration of the PEG molecules per surface area unit we need to calculate the total surface ($SA_{total}$) area. As a first step we calculate the surface area of a single nanoparticle ($SA_{NP}$):

$$SA_{NP} = 4\pi R^2, \tag{S3}$$

where $R$ is the radius in nm (taken from DCS analysis). The number of nanoparticles can be calculated using the concentration of silica particles, the volume of a sphere and the material density as:

$$N_{NP} = \frac{3m 10^{18}}{4\pi d r^3}, \tag{S4}$$

where $m$ is the mass of the silica sample (measured through vaccum drying of aliquots of known volume) and $d$ is silica density (taken as 2.0 g/cm³).

Combining Eqs. 3 and 4 we obtain for the total surface area:

$$Total\ SA = \frac{3m 10^{18}}{4\pi d\ R^3} \cdot 4\pi R^2 = \frac{3m 10^{18}}{d\ R}, \tag{S5}$$

We can obtain the number of PEG molecules directly from our NMR results. When we combine Eq. 5 with the NMR results we obtain for the PEG concentration on the surface:

$$\frac{PEG}{nm^2} = \frac{N_{PEG}}{Total\ SA} = \frac{1}{3}\frac{d\ R N_{PEG}}{m} 10^{-18}, \tag{S6}$$

where $\frac{PEG}{nm^2}$ is the concentration of PEG on the particle surface, $N_{PEG}$ is the concentration of PEG in the dissolved particle solution measured by NMR

**NFM Measurement**

*Optofluidics instrument*

The Optofluidics NanoTweezer instrumental setup consists of an instrument which contains a laser (635 mW, 1064 nm), a pneumatic pump (can regulate from 0 to 70 mBar of pressure) for fluid flow and additional electronic and optics. The instrument is linked to a microchip mounted on a microscope stage and the microscope was further equipped with a camera (figure S1).

Each Optofluidics NanoTweezer chip contains five silicon nitride ($Si_3N_4$) waveguides (WG) two 1, two 1.5 and one 2 μm wide which is situated in a 200 μm x 200 μm experimental chamber. The rest of the chips is covered by $SiO_2$ (~8 μm thick outside the WG chamber and several nm in the chamber) (figure S2). 1064 nm laser light (TE mode polarized) is supplied by the instrument laser, coupled to the waveguides by the pre-aligned optical fibers, and guided to the waveguide outputs where optical power is measured with a photodiode. A sample is introduced by inserting an aspirator into the solution of interest. The sample is drawn through the system with vacuum pressure and ultimately collected in a waste reservoir. Vacuum pressure is regulated in the range of 0 to 70 mBar, and can be increased to ~300 mBar for rapid sample loading and washing. Precise flow rate control in the range of 0 - 7 μL/min is achieved by using an in-line flow rate sensor and a PID feedback control loop. This movement enables steady-state imaging of particles as they travel along the waveguide and pass through the imaging region of interest. The intense scattering generated by particles enables high signal to noise imaging at low (100 μs) exposure time and high frame rates 2555 fps. Images recorded by the camera are analyzed with a custom software package that performs automated particle tracking.

Each WG is coupled to optical fibers at both its ends, as the laser light exits the WG the optical power is measured with a photodiode in order to monitor the optical properties of the chip. The sample is loaded onto the chip using the pneumatic pump at increased pressure (~300 mBar).

Waveguides were imaged with an upright transmission light microscopy (Olympus BX60, Japan). The system was firstly run with deionized water for 5 min, and then the waveguides were evaluated with a dry 20 X objective lens (0.40 NA) for any presence of debris. The inlet tube was placed into a desired sample and purged at a pressure rate of ~300 mBar until the chip was entirely filled. A lower laser power was firstly applied (50 mW) to make sure there was no bubble on the WG. Before a movie was taken the power was set to what was required and flow to 0.1 µL/min. Trapping objects on a waveguide were focused with an Olympus LUCPLFLN40X objective lens (0.6 NA) and images were captured for 20 seconds using a Basler acA2000-165uc camera. For each particle type, between 15 and 25 movies were acquired depending on sample behavior. In between measurement of different particle types, sufficient amount of deionized water was run through the waveguides to avoid cross-contamination. In some rare cases of tough contamination ethanol was flowed through the system.

*Trajectory selection*

It was found that a further trajectory selection procedure is required after the movie was analysed. The same movie analysis conditions were used thought this study which in many cases led to "artefact trajectories" which needed to be removed manually. We plotted and considered the intensity over time, place of the trajectory on the WG, the diffusion along the X and Y axis and the total potential for each particle (figure S9). Using this information we can exclude many of the artefacts observed.

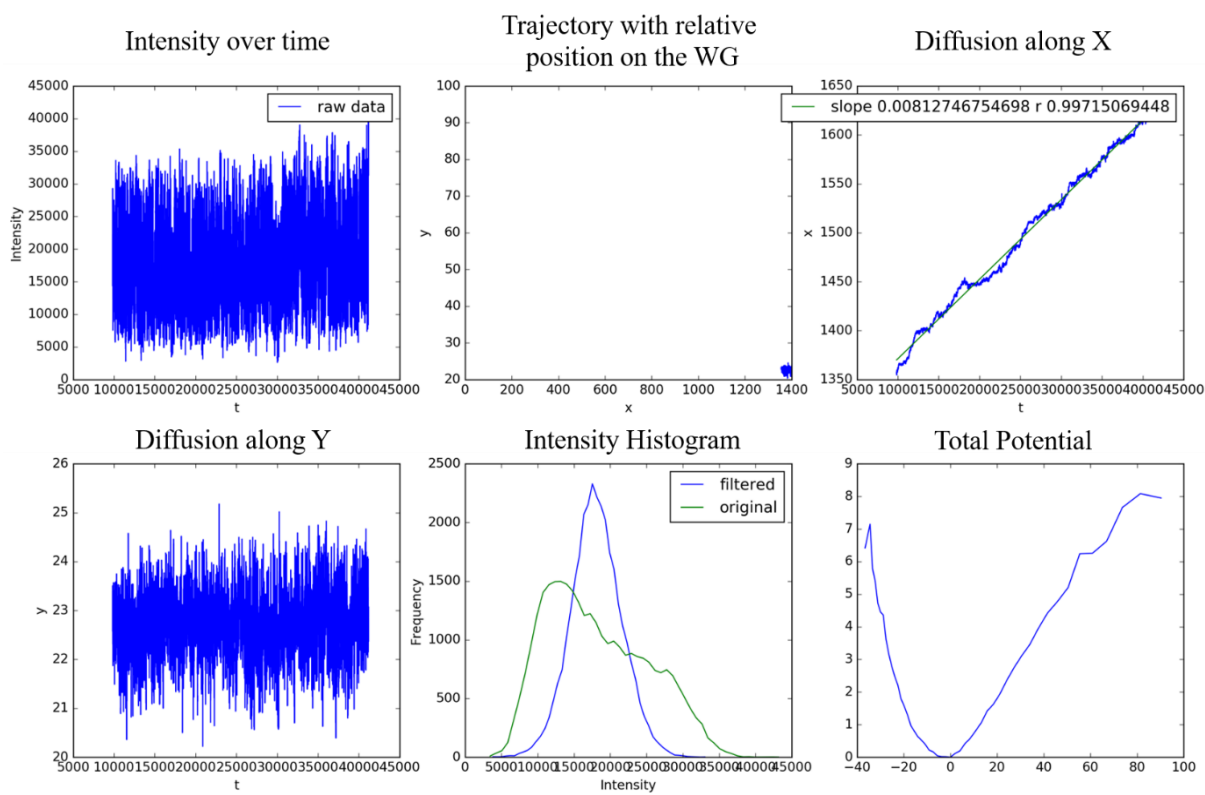

**Figure S9.** Example information sheet acquired for each particle trajectory. It includes the change of particle intensity over time, its movement in the X and Y axis, the relative position on the WG, the particle intensity histogram and total potential.

Examples of trajectories which were considered for analysis are presented in Figure S10a and S10b. Others which were excluded manually include tracks at which the particle is there only for a small fraction of the total frames (Figure S10c and S10d). This type of artefact is characterized by a lack of mobility on either axis and an intensity several times lower than a typical particle trajectory. The particle may interact with features of the WG surface or get stuck for parts of its trajectory this leads to abnormal behavior (figure S10e and f). Such trajectories typically have a very distinguishable total potential. In some cases the Airy disk of the particle may be mistaken for another particle. Similarly to background, these trajectories are defined with a very low intensity and move together with another trajectory.

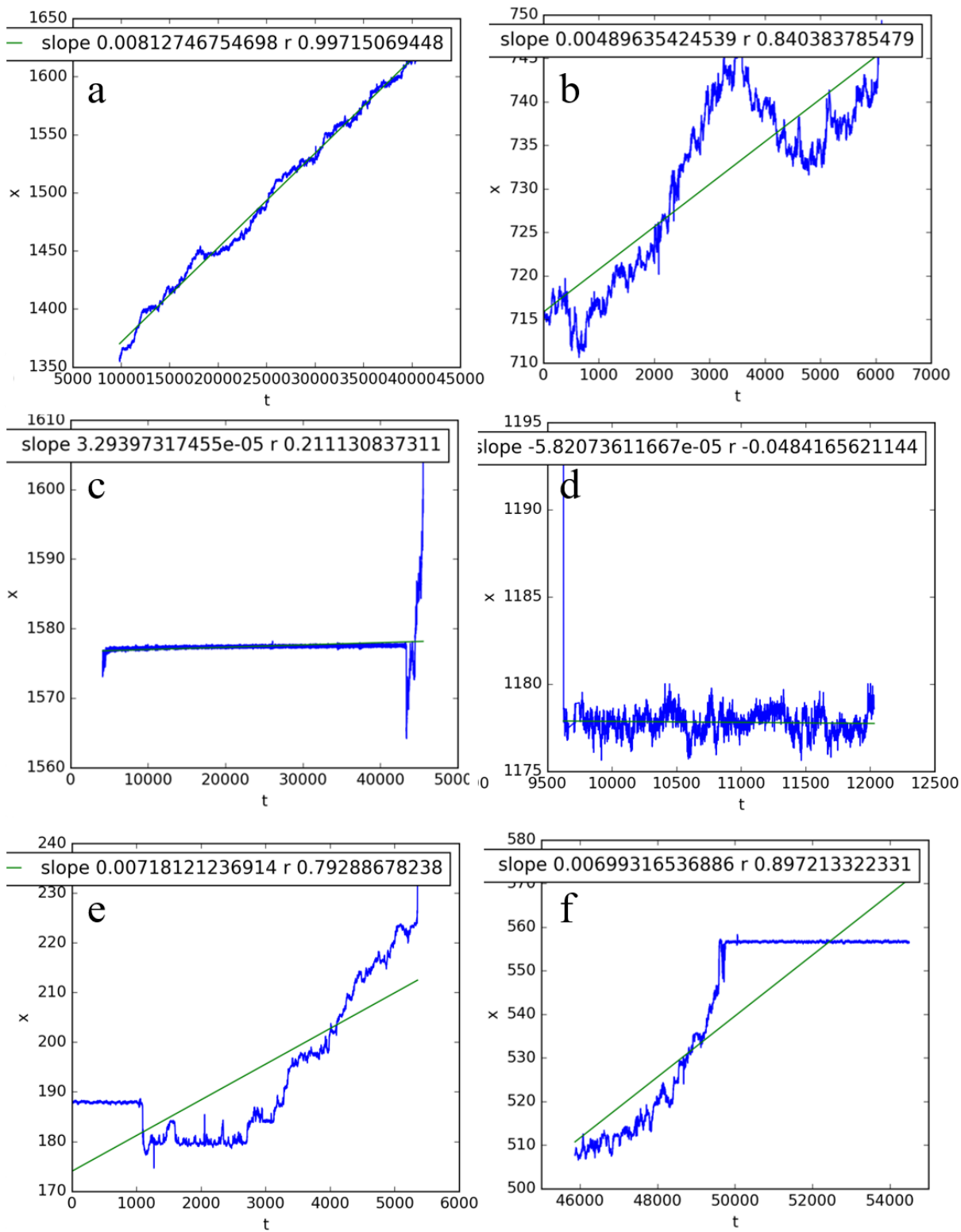

**Figure S10.** Examples of different trajectories commonly found in the NFM experiment. Those include examples of a (a) perfect trajectory and (b) an acceptable trajectory as well as examples of artefacts such as background (c) with and (d) without a particle appearing in the trajectory.